\documentclass[aps,prc,twocolumn ,superscriptaddress]{revtex4-1}

\usepackage{graphicx}
\usepackage{amsmath}
\usepackage{braket}

\usepackage[colorlinks,
linkcolor=blue,
anchorcolor=blue,
urlcolor=blue,
citecolor=blue]{hyperref}

\begin{document}


\title{\textit{Ab-initio} no-core  Gamow shell model calculations of multi-neutron systems}


\author{J.G. Li}
\affiliation{School of Physics,  and   State Key  Laboratory  of  Nuclear  Physics   and  Technology, Peking University, Beijing  100871, China}
\author{N. Michel}
\affiliation{Institute of Modern Physics, Chinese Academy of Sciences, Lanzhou 730000, China}
\affiliation{School of Nuclear Science and Technology, University of Chinese Academy of Sciences, Beijing 100049, China}
\author{B.S. Hu}
\affiliation{School of Physics,  and   State Key  Laboratory  of  Nuclear  Physics   and  Technology, Peking University, Beijing  100871, China}
\author{W. Zuo}
\affiliation{Institute of Modern Physics, Chinese Academy of Sciences, Lanzhou 730000, China}
\affiliation{School of Nuclear Science and Technology, University of Chinese Academy of Sciences, Beijing 100049, China}
\author{F.R. Xu}\email[]{frxu@pku.edu.cn}
\affiliation{School of Physics,  and   State Key  Laboratory  of  Nuclear  Physics   and  Technology, Peking University, Beijing  100871, China}



\date{\today}

\begin{abstract}

 The existence of multi-neutron systems has always been a debatable question. Indeed, both inter-nucleon correlations and a large continuum coupling occur in these states. We then employ the \textit{ab-initio} no-core  Gamow shell model to  calculate the resonant energies and widths of the trineutron  and tetraneutron systems with realistic interactions. Our results indicate that trineutron and tetraneutron are both unbound and bear broad widths. The calculated energy and width of tetraneutron are also comparable with recent experimental data. Moreover, our calculations  suggest that the energy of trineutron is lower than that of tetraneutron, while its resonance width is also narrower. This strongly suggests that trineutron is more likely to be experimentally observed than tetraneutron. We thus suggest experimentalists to search for trineutron at low energy.

\end{abstract}


\maketitle

 \section{Introduction}   
 Few-body multi-neutron systems are located close to the neutron drip line. Thus, they provide us with a unique laboratory  to  understand  nuclear properties  at drip lines \cite{PhysRevLett.55.2676,PhysRevLett.116.102503,PhysRevLett.120.152504} and nuclear forces in the absence of Coulomb interaction \cite{PhysRevC.68.041001}.

 Tremendous efforts have been made during the last few decades to understand few-nucleon systems, in particular few-neutron systems \cite{History_multi-neutron,PhysRevLett.116.052501,PhysRevC.65.044006}.  Earlier experiments failed to find positive evidence for the existence of multi-neutron systems (see a short summary in Ref.\cite{History_multi-neutron}).  In 2002, Marqu{\'e}s \textit{et al.} reported that a bound tetraneutron  was observed in a breakup reaction of the $^{14}$Be $\to$ $^{10}$Be+4n channel \cite{PhysRevC.65.044006}. After that experiment, several theoretical attempts were performed to examine the possible existence of bound tetraneutron, but all the calculations failed to reproduce experimental data \cite{PhysRevLett.90.252501,Bertulani_2003}.  The interest in multi-neutron systems has been resurrected through an experiment where a candidate for resonant tetraneutron was observed, whose resonant energy and width were measured in the doubly charge-exchange reaction $^{4}$He($^8$He,$^8$Be) \cite{PhysRevLett.116.052501}: $E_r  = 0.83 \pm 0.65(stat) \pm 1.25(syst)$ MeV and $\Gamma$ $\leq$ 2.6 MeV, respectively. However, they happen to bear large error bars, so that the measurements are compatible with a bound tetraneutron.  Several other experiments were then approved to reconsider measurements related to tetraneutron \cite{tetra_neutron_1,tetra_neutron_2,tetra_neutron_3}.

 There were numerous theoretical calculations taking into account inter-nucleon correlations and high-quality realistic nucleon-nucleon interactions in the frame of \textit{ab-initio} frameworks in recent years \cite{PhysRevLett.90.252501,Bertulani_2003,PhysRevLett.117.182502,PhysRevLett.118.232501,PhysRevLett.119.032501,PhysRevC.97.034001,DELTUVA2018238,PhysRevC.72.034003,PhysRevC.93.044004}.
 The extrapolation of  no-core shell model (NCSM) calculations utilizing the realistic two-body JISP16 interaction with a harmonic oscillator (HO) basis showed that tetraneutron is located near 0.8 MeV above threshold and that its width  is about  1.4 MeV  \cite{PhysRevLett.117.182502}. Recently, NCSM calculations performed in larger model spaces  provided with two resonant states of tetraneutron, at 0.3 MeV and 0.8 MeV, of widths about 0.85 MeV and 1.3 MeV, respectively \cite{Tetraneutron_NCSM2018}.
 Calculations have also been done in the frame of  no-core Gamow shell model (NCGSM), using the density matrix renormalization group (DMRG) method and natural orbitals (n.o.) \cite{PhysRevLett.119.032501}. It was speculated that the resonance width of tetraneutron is larger than 3.7 MeV, so that the formation of a nucleus therein is precluded. However, the calculations were incomplete, as they could be performed only in truncated model spaces or with unphysically overbinding interactions,
 so that no definite conclusion could be made \cite{PhysRevLett.119.032501}.

 The quantum Monte Carlo (QMC) framework has been used to calculate multi-neutron systems based on local chiral interactions via an extrapolation of bound states (obtained with a confining auxiliary potential) to the physical domain of unbound multi-neutron systems.  The calculations predicted that the trineutron is located at 1.11 MeV, below the energy of tetraneutron, at 2.12 MeV \cite{PhysRevLett.118.232501,PhysRevLett.123.069202}.
The Faddeev method has also been employed for that matter \cite{PhysRevC.93.044004,PhysRevC.72.034003,PhysRevC.97.034001,DELTUVA2018238}. The latter calculations suggested that the existence of multi-neutron resonances can only be obtained by strongly modifying standard nuclear forces, which is incompatible with our current understanding of nucleon-nucleon interactions \cite{PhysRevC.93.044004}.  The calculations of Ref.\cite{PhysRevLett.119.032501,PhysRevLett.118.232501,PhysRevLett.117.182502} showed that multi-neutron systems are sensitive neither to various realistic interactions, in the presence or absence of three-body forces,  nor to the momentum cutoff present in renormalization group methods. In this paper, we will then present a nearly exact study of trineutron and tetraneutron systems employing \textit{ab-initio}  NCGSM \cite{PhysRevLett.119.032501,PhysRevC.88.044318}. Indeed, NCGSM comprises both a consistent treatment of many-body correlations and coupling to the continuum by using the Berggren ensemble \cite{BERGGREN1968265}. Both the resonant energies and widths of the trineutron and tetraneutron systems will be presented.

\section{The Method} 
 A resonance state, which represents a decaying process, is time-dependent. However, the exact treatment of time-dependence is difficult to handle theoretically in many-body system (see Ref.\cite{PhysRevC.79.044308} for examples of time-dependent shell model calculations in light nuclei).
In order to be able to consider many-body resonance states in a time-independent framework, it is convenient to reformulate the  Schr$\ddot{\rm o}$dinger  equation in the complex momentum plane. For this, one uses the Berggren basis, comprising bound, resonant and scattering single-particle (s.p.) states. The time dependence is then taken into account by removing the hermitian character of the Hamiltonian, so that eigenvalues become complex. The imaginary part of the energy is indeed proportional to the decay width \cite{BERGGREN1968265}.

The completeness relation, introduced by Berggren \cite{BERGGREN1968265}, reads,
\begin{eqnarray}\label{equation2}
   \sum_{n} |\phi_{nj}\rangle\langle \widetilde {\phi}_{nj}| + \frac{1}{\pi} \int_{L_{+}}|\phi_{j} (k)\rangle\langle \phi_{j} (k^{\ast})| dk = 1,
\end{eqnarray}
where $\phi_{nj}$ are the Berggren-basis pole states of bound and decaying character, and $\phi_{j} (k)$ is the scattering state belonging to the $L_{+}$ contour of complex momenta. In practical calculations, the integral in Eq.(\ref{equation2}) is discretized by means of an appropriate quadrature rule (the Gauss-Legendre quadrature in our case).

In the present work, we employ the NCGSM method to calculate multi-neutron systems. Many-body states  are the linear combinations of the Slater determinants $|SD_{n}\rangle = |u_1,....,u_A\rangle$, where $u_k$ is a bound, resonant or non-resonant (scattering) state. The Hamiltonian matrix in the NCGSM is complex symmetric and bears complex eigenvalues. Coupling to the continuum is then present at basis level, and many-body correlations occur through configuration mixing \cite{PhysRevLett.89.042502,PhysRevLett.89.042501,PhysRevLett.119.032501,BARRETT2013131,PhysRevC.88.044318}.  The widths of the eigenstates obtained in NCGSM
take all particle-emission channels into account, so that they are total decay widths.
Conversely, other methods, as those using Faddeev-Yakubovsky equations \cite{DELTUVA2018238} and NCSM calculations \cite{PhysRevLett.117.182502},  need to consider all partial decay processes to calculate the total decay width.

The initial realistic Hamiltonian reads,
\begin{eqnarray}\label{equation1}
    H  = \frac{1}{A} \sum_{i}^{A} \frac{(p_i-p_j)^2}{2m} + \sum_{i<j}^{A} V^{i<j}_{NN}
\end{eqnarray}
where $V^{i<j}_{NN}$ is a realistic two-body nucleon-nucleon interaction.

We are looking for the NCGSM eigenstates which give the resonant multi-neutron states in physics.
In the NCGSM calculations, the lowest eigenvalue is unphysical in general. Indeed, it is usually a many-body scattering state. To find the physical lowest state from the scattering states in NCGSM calculations, we use the overlap method \cite{Michel_2008}. It consists in identifying physical  states by searching for the eigenstates which bear the largest overlap with those arising from the pole approximation \cite{Michel_2008} in which the valence configurations are built from the pole space of the Berggren basis \cite{Michel_2008}. The Lanczos method is indeed sufficient to calculate eigenstates at the pole approximation level.
However, the overlap method based on the pole approximation is insufficient for multi-neutron systems due to the extremely strong coupling to the scattering basis states.

In order to be able to identify physical resonant states of multi-neutron systems, we added an auxiliary Wood-Saxon (WS) potential
to the Hamiltonian of Eq.(\ref{equation1}). It reads $V_{\textrm {WS-aux}}(r) = V_{\rm {aux}}/[1+e^{(r-R_{\rm{aux}})/a_{\rm{aux}}}]$, where the depth $V_{\rm {aux}}$ will vary from a finite value to zero, the radius $R_{\rm {aux}}$ is fixed at 4 fm and the diffuseness $a_{\rm {aux}}$ is equal to 0.65 fm. The auxiliary WS potential thus confines the valence neutrons in an external trap. This method has  been employed in the QMC framework \cite{PhysRevLett.90.252501,PhysRevLett.118.232501} to extrapolate the energies of the multi-neutron systems from negative energies to the physical region. The precision of obtained extrapolation has been checked to be nearly exact for the dineutron system \cite{PhysRevLett.123.069202}. In the present calculations, both resonance and continuum channels are included, and hence we can calculate the energies of multi-neutron systems in the physical region and give the resonance widths of multi-neutron systems. The convergences of model spaces have been also well tested to obtain precise results. 
We checked as well that no new particle-emission channel opens when decreasing $V_{\rm aux}$, so that the used method for  $V_{\rm aux} \rightarrow $ 0 is justified.

We firstly obtain many-body bound states of multi-neutron systems using the auxiliary WS potential.
Then, $V_{\rm aux}$ is slowly diminished, so that the former states can be used as reference states in the overlap method of Ref.\cite{Michel_2008}.
One can then determine the physical multi-neutron systems by iteratively using the previously calculated many-body eigenstates as reference states,
smoothly decreasing $V_{\rm aux}$ until $V_{\rm aux} = 0$. A similar technique was used in the NCGSM calculations of Ref.\cite{PhysRevLett.119.032501} to identify broad resonance in tetraneutron, where overbinding interactions were used instead of a confining auxiliary WS potential.

\section{Calculations and discussions} 
In the present work, the s.p.~Berggren basis is generated by  a finite-depth WS potential including spin-orbit coupling. The parameters of the WS potential read $R_0$ = 1.9 fm for its radius, $a$ = 0.67 fm for the diffuseness, $V_{ls}$ = 9.5 MeV for its spin-orbit strength and $V_0 = -27$ MeV for the central depth. With those parameters, the $0s_{1/2} $ and $0p_{3/2}$ orbitals are bound states, and the non-resonant scattering partial waves consist of  the $L_{+}$ contours (in the complex momentum space) defined by the coordinate points $(0,0),(0.25,-0.24),(0.5,0.0)$ and $(4.0,0.0)$ (all in fm$^{-1}$). Each segment of the contours $L_+$ is discretized with 15 points.  We employ the chiral two-body interaction N$^3$LO \cite{PhysRevC.68.041001}, renormalized by the way of the $V_{ {\rm low}\textrm{-}k}$ method using a momentum cutoff $\Lambda$ = 2.1 fm$^{-1}$. The results for multi-neutron systems have been checked to be sensitive neither to the realistic nuclear forces used nor to the cutoff of the renormalization method \cite{PhysRevLett.119.032501,PhysRevLett.118.232501}. This is partly due to the dilute density of multi-neutron systems, as  noticed in Refs.\cite{PhysRevLett.118.232501,PhysRevLett.119.032501}. We have also checked the basis-dependence by changing $V_{\rm{0}}$ for trineutron. The trineutron energy varies by only about 50 keV if $V_0$ changes by a few MeV. Therefore, basis dependence is negligible in our calculations.
We take a model space composing of the $s_{1/2}$ and $p_{3/2}$ partial waves using the Berggren basis, while the other partial waves, i.e.~$p_{1/2},d,f$ and $g_{9/2}$, are represented by HO basis states. All HO orbitals satisfy $N_{max} = 2n+l \le  20 $. The oscillator length for the HO basis is $b$ = 2 fm. It has been checked in Ref.\cite{PhysRevLett.119.032501} that calculations are almost independent of the length parameter and of the inclusion of other partial waves.

\begin{figure}[!htb]
\includegraphics[width=1.08\columnwidth]{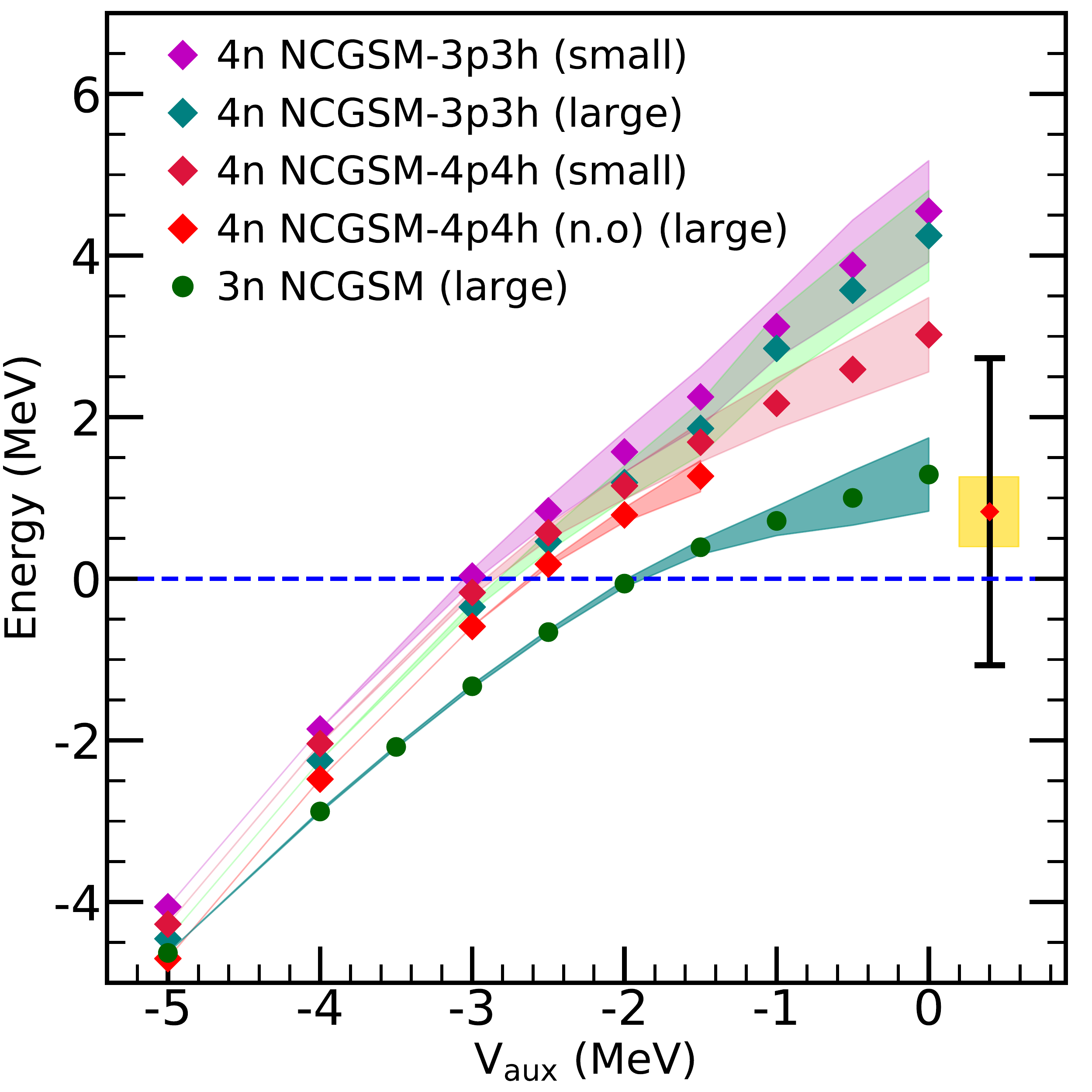}
\caption{Evolution of the energies and widths of trineutron and tetraneutron as function of the depth of the auxiliary WS potential in the small and large spaces with NCGSM-3p3h and NCGSM-4p4h truncations, using either the Berggren or n.o.~basis (see text for definitions). n.o.~has been used in NCGSM-4p4h space for $V_{\rm aux} < -1.5$ MeV as calculations become unstable afterwards. The width of tetraneutron has been decreased by a factor 3 for readability on the figure. Trineutron results have been obtained without truncations. The experimental datum \cite{PhysRevLett.116.052501} for tetraneutron is also shown.}\label{fig:1}
\end{figure}

A complete diagonalization of the Hamitonian for trineutron can be done using the model space described above using NCGSM. However, it is not possible for tetraneutron because of the huge model  dimension and strong coupling to continuum basis states. The full model space is then called the large space.
Calculations in NCGSM-3p3h truncated model spaces, i.e.~with three neutrons occupying scattering basis states at most, can be done. Thus, in order to obtain nearly exact (NCGSM-4p4h) results for tetraneutron, we first did calculations in a model space bearing fewer high lying orbitals, defined by taking all the $s_{1/2}$ and $p_{3/2}$ Berggren basis states, two HO shells for $p_{1/2},d$ and one HO shell $0f$ and $0g_{9/2}$ for the remaining partial waves. Calculations without truncations can then be effected in this model space. This model is deemed as the small space. By comparing the results obtained with NCGSM-3p3h and NCGSM-4p4h in the small and large spaces, using n.o.~with $V_{\rm{aux}} <-1.5$ MeV with NCGSM-4p4h in the large space, we could get an estimate of the error made on energy and width for tetraneutron (see Fig. \ref{fig:1}). We checked that the use of n.o.~provides virtually the same results as the Berggren basis in small spaces using NCGSM-4p4h.
Note that n.o.~could not lead to stable results for $V_{\rm{aux}} \simeq 0$ MeV due to the very large width of tetraneutron, a phenomenon also present in Ref.\cite{PhysRevLett.119.032501}. We notice that energy is about 360 keV too unbound in the small space with either NCGSM-3p3h or NCGSM-4p4h. We also remark that the width is about 380 keV too broad in the unbound region of tetraneutron. These estimates are almost independent of $V_{\rm{aux}}$ . Consequently, in order to estimate the energy and width of tetraneutron in the large space, we calculate the energy and width of tetraneutron in the small space without truncations and remove 360 keV for energy and 380 keV for width from the latter values, respectively.  We can then assume that the exact energy and width are nearly the same as with this ansatz.

For  the four-neutron system, when the strength of the confining WS potential gradually weakens, the widths of resonances increase quickly. Moreover, the differences of energies using NCGSM-3p3h and NCGSM-4p4h model spaces also become gradually larger therein. These results suggest that, in general, resonances bear a strong coupling to the non-resonant continuum, with the induced configuration mixing becoming more and more important when the width increases. Thus, coupling to continuum should be considered properly in the treatment of nuclei close to drip-line \cite{PhysRevLett.89.042502,PhysRevLett.117.222501,SUN2017227,PhysRevLett.108.242501}.

\begin{figure}[!htb]
\includegraphics[width=1.00\columnwidth]{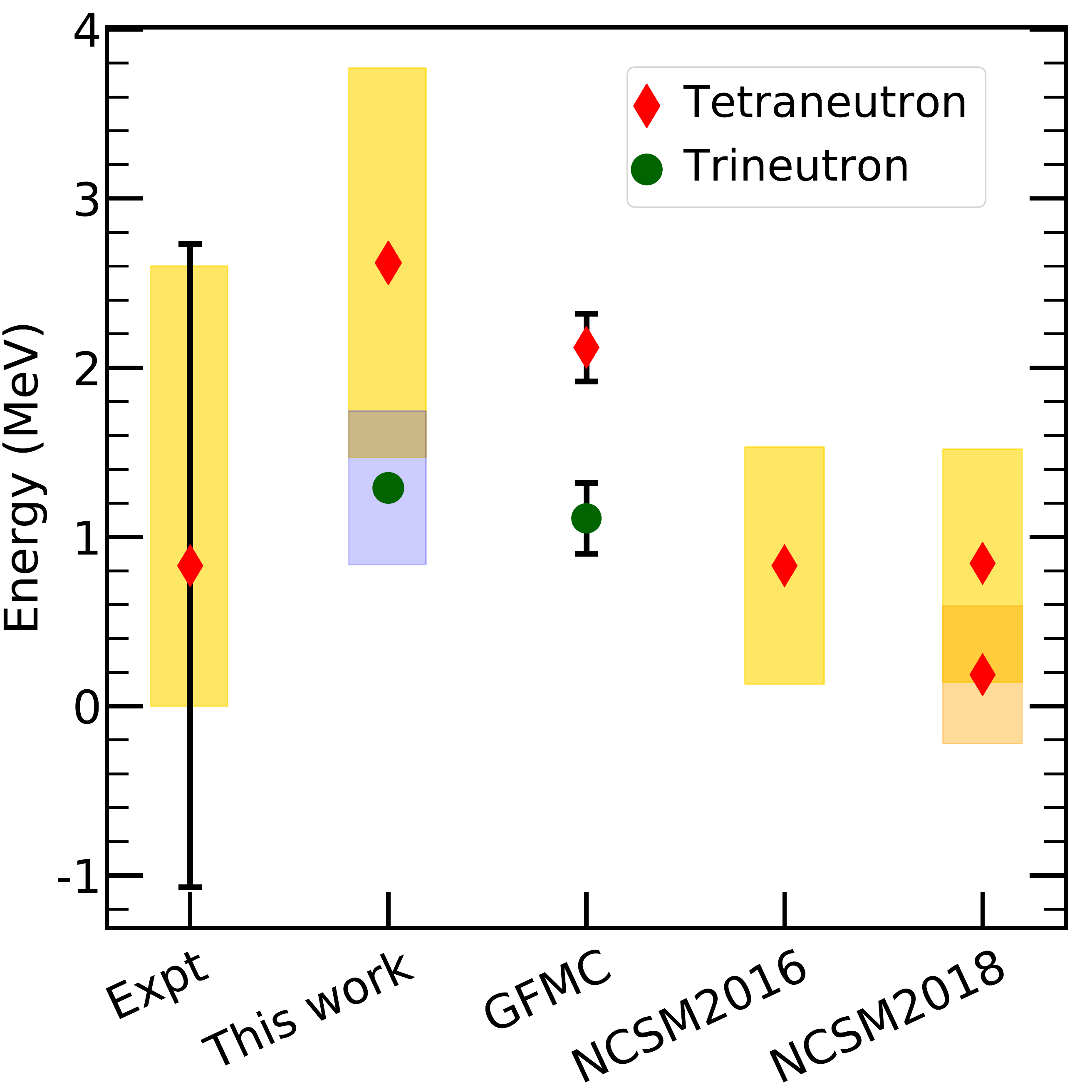}
\caption{Calculated energies of trineutron and tetraneutron by different models. The corresponding references are QMC: \cite{PhysRevLett.118.232501}, NCSM2016: \cite{PhysRevLett.117.182502} and NCSM2018: \cite{Tetraneutron_NCSM2018}. The experimental datum is from Ref.\cite{PhysRevLett.116.052501}.}\label{fig:2}
\end{figure}

The final results for the three-neutron and four-neutron systems, i.e.,~for which $V_{\rm aux} = 0$, and with tetraneutron energies and widths modified by 360 keV  and 380 keV, respectively, following the method explained above, as well as experimental data for tetraneutron, are shown in Fig. \ref{fig:2}.  Our calculations provide with neutron-unbound trineutron and tetraneutron, whose energies are 1.29 MeV and 2.64 MeV, respectively. We also predict the widths of trineutron and tetraneutron to be  $\Gamma_{3n}$  = 0.91 MeV and $\Gamma_{4n}$ = 2.38 MeV. The calculated energy and width of tetraneutron are within the range of the experimental error \cite{PhysRevLett.116.052501}. Our calculations predict the energy of the trineutron to be lower than that of tetraneutron and that the width of the trineutron is smaller than that of tetraneutron. Consequently, we suppose that trineutron should be more likely to be  observed experimentally than tetraneutron.
We also compared our calculations with other \textit{ab-initio} predictions for multi-neutron systems, see Fig.\ref{fig:2}.  The  energies of  trineutron and tetraneutron  in our calculations are close to the  extrapolated results of the QMC calculations \cite{PhysRevLett.118.232501}.
Moreover, our calculated energy for tetraneutron is higher than that of NCSM calculations, which is clearly due to the neglect of continuum coupling in NCSM.
The present prediction of energy for tetraneutron is similar to that of K.~Fossez \textit{et al.} \cite{PhysRevLett.119.032501}. Indeed, the extrapolation of energy arising from the DMRG calculation in Ref.\cite{PhysRevLett.119.032501} (where no truncations are made) provides more or less with the same tetraneutron energy as in our calculations. However, the predicted width for tetraneutron of Ref.\cite{PhysRevLett.119.032501} is not consistent with ours.
It was speculated therein \cite{PhysRevLett.119.032501} that the width of tetraneutron is larger than 3.7 MeV, whereas our calculation predicts a width of about 2.38 MeV.
We explain this discrepancy from the fact that the predicted width from K.~Fossez \textit{et al.} \cite{PhysRevLett.119.032501} was based on a calculation effected in a NCGSM-2p2h truncated space.
Our present calculations have shown that the calculated width of tetraneutron decreases by considering more correlations, i.e.,~by using the NCGSM-3p3h and NCGSM-4p4h spaces.

\section {Summary} 

 In this paper, we utilized the $ab$-$initio$ NCGSM framework along with the realistic chiral two-body  N$^3$LO nuclear force to calculate  few-body multi-neutron systems. In the NCGSM framework, inter-nucleon correlations and continuum coupling are taken into account. Our calculations predict both trineutron and tetraneutron systems to be broad resonances, with energies equal to  1.29 and 2.64 MeV, and  widths equal to  0.91 and 2.38 MeV, respectively. The calculated energy and width of tetraneutron are both within the range of the experimental error \cite{PhysRevLett.116.052501}.
Our results also show that the energy  of trineutron is lower than that of tetraneutron, and widths present a similar tendency.  As trineutron has smaller energy and width than tetraneutron,   one can suppose that trineutron would more likely to be observed than tetraneutron.  We then encourage more experiments to be carried out to study the trineutron resonance state.

\begin{acknowledgments}
 Valuable discussions with W. Nazarewicz, M. P{\l}oszajczak, Dean Lee,  J. P. Vary, Z.H. Yang, S.M. Wang, J.C. Pei and G.C. Yong are gratefully acknowledged.
  This work has been supported by the National Key R\&D Program of China under Grant No. 2018YFA0404401; the National Natural Science Foundation of China under Grants No. 11835001, No. 11921006, No. 11575007, No. 11847203, No. 11435014 and No. 11975282; China Postdoctoral Science Foundation under Grant No. 2018M630018; and the CUSTIPEN (China-U.S. Theory Institute for Physics with Exotic Nuclei) funded by the U.S. Department of Energy, Office of Science under Grant No. de-sc0009971. We acknowledge the High-Performance Computing Platform of Peking University for providing computational resources.
\end{acknowledgments}

\bibliography{Multi-neutron_systems}

\begin{thebibliography}{31}%
\makeatletter
\providecommand \@ifxundefined [1]{%
 \@ifx{#1\undefined}
}%
\providecommand \@ifnum [1]{%
 \ifnum #1\expandafter \@firstoftwo
 \else \expandafter \@secondoftwo
 \fi
}%
\providecommand \@ifx [1]{%
 \ifx #1\expandafter \@firstoftwo
 \else \expandafter \@secondoftwo
 \fi
}%
\providecommand \natexlab [1]{#1}%
\providecommand \enquote  [1]{``#1''}%
\providecommand \bibnamefont  [1]{#1}%
\providecommand \bibfnamefont [1]{#1}%
\providecommand \citenamefont [1]{#1}%
\providecommand \href@noop [0]{\@secondoftwo}%
\providecommand \href [0]{\begingroup \@sanitize@url \@href}%
\providecommand \@href[1]{\@@startlink{#1}\@@href}%
\providecommand \@@href[1]{\endgroup#1\@@endlink}%
\providecommand \@sanitize@url [0]{\catcode `\\12\catcode `\$12\catcode
  `\&12\catcode `\#12\catcode `\^12\catcode `\_12\catcode `\%12\relax}%
\providecommand \@@startlink[1]{}%
\providecommand \@@endlink[0]{}%
\providecommand \url  [0]{\begingroup\@sanitize@url \@url }%
\providecommand \@url [1]{\endgroup\@href {#1}{\urlprefix }}%
\providecommand \urlprefix  [0]{URL }%
\providecommand \Eprint [0]{\href }%
\providecommand \doibase [0]{http://dx.doi.org/}%
\providecommand \selectlanguage [0]{\@gobble}%
\providecommand \bibinfo  [0]{\@secondoftwo}%
\providecommand \bibfield  [0]{\@secondoftwo}%
\providecommand \translation [1]{[#1]}%
\providecommand \BibitemOpen [0]{}%
\providecommand \bibitemStop [0]{}%
\providecommand \bibitemNoStop [0]{.\EOS\space}%
\providecommand \EOS [0]{\spacefactor3000\relax}%
\providecommand \BibitemShut  [1]{\csname bibitem#1\endcsname}%
\let\auto@bib@innerbib\@empty
\bibitem [{\citenamefont {Tanihata}\ \emph {et~al.}(1985)\citenamefont
  {Tanihata}, \citenamefont {Hamagaki}, \citenamefont {Hashimoto},
  \citenamefont {Shida}, \citenamefont {Yoshikawa}, \citenamefont {Sugimoto},
  \citenamefont {Yamakawa}, \citenamefont {Kobayashi},\ and\ \citenamefont
  {Takahashi}}]{PhysRevLett.55.2676}%
  \BibitemOpen
  \bibfield  {author} {\bibinfo {author} {\bibfnamefont {I.}~\bibnamefont
  {Tanihata}}, \bibinfo {author} {\bibfnamefont {H.}~\bibnamefont {Hamagaki}},
  \bibinfo {author} {\bibfnamefont {O.}~\bibnamefont {Hashimoto}}, \bibinfo
  {author} {\bibfnamefont {Y.}~\bibnamefont {Shida}}, \bibinfo {author}
  {\bibfnamefont {N.}~\bibnamefont {Yoshikawa}}, \bibinfo {author}
  {\bibfnamefont {K.}~\bibnamefont {Sugimoto}}, \bibinfo {author}
  {\bibfnamefont {O.}~\bibnamefont {Yamakawa}}, \bibinfo {author}
  {\bibfnamefont {T.}~\bibnamefont {Kobayashi}}, \ and\ \bibinfo {author}
  {\bibfnamefont {N.}~\bibnamefont {Takahashi}},\ }\href {\doibase
  10.1103/PhysRevLett.55.2676} {\bibfield  {journal} {\bibinfo  {journal}
  {Phys. Rev. Lett.}\ }\textbf {\bibinfo {volume} {55}},\ \bibinfo {pages}
  {2676} (\bibinfo {year} {1985})}\BibitemShut {NoStop}%
\bibitem [{\citenamefont {Kondo}\ \emph {et~al.}(2016)\citenamefont {Kondo},
  \citenamefont {Nakamura}, \citenamefont {Tanaka}, \citenamefont {Minakata},
  \citenamefont {Ogoshi}, \citenamefont {Orr}, \citenamefont {Achouri},
  \citenamefont {Aumann}, \citenamefont {Baba}, \citenamefont {Delaunay},
  \citenamefont {Doornenbal}, \citenamefont {Fukuda}, \citenamefont {Gibelin},
  \citenamefont {Hwang}, \citenamefont {Inabe}, \citenamefont {Isobe},
  \citenamefont {Kameda}, \citenamefont {Kanno}, \citenamefont {Kim},
  \citenamefont {Kobayashi}, \citenamefont {Kobayashi}, \citenamefont {Kubo},
  \citenamefont {Leblond}, \citenamefont {Lee}, \citenamefont {Marqu\'es},
  \citenamefont {Motobayashi}, \citenamefont {Murai}, \citenamefont {Murakami},
  \citenamefont {Muto}, \citenamefont {Nakashima}, \citenamefont {Nakatsuka},
  \citenamefont {Navin}, \citenamefont {Nishi}, \citenamefont {Otsu},
  \citenamefont {Sato}, \citenamefont {Satou}, \citenamefont {Shimizu},
  \citenamefont {Suzuki}, \citenamefont {Takahashi}, \citenamefont {Takeda},
  \citenamefont {Takeuchi}, \citenamefont {Togano}, \citenamefont {Tuff},
  \citenamefont {Vandebrouck},\ and\ \citenamefont
  {Yoneda}}]{PhysRevLett.116.102503}%
  \BibitemOpen
  \bibfield  {author} {\bibinfo {author} {\bibfnamefont {Y.}~\bibnamefont
  {Kondo}}, \bibinfo {author} {\bibfnamefont {T.}~\bibnamefont {Nakamura}},
  \bibinfo {author} {\bibfnamefont {R.}~\bibnamefont {Tanaka}}, \bibinfo
  {author} {\bibfnamefont {R.}~\bibnamefont {Minakata}}, \bibinfo {author}
  {\bibfnamefont {S.}~\bibnamefont {Ogoshi}}, \bibinfo {author} {\bibfnamefont
  {N.~A.}\ \bibnamefont {Orr}}, \bibinfo {author} {\bibfnamefont {N.~L.}\
  \bibnamefont {Achouri}}, \bibinfo {author} {\bibfnamefont {T.}~\bibnamefont
  {Aumann}}, \bibinfo {author} {\bibfnamefont {H.}~\bibnamefont {Baba}},
  \bibinfo {author} {\bibfnamefont {F.}~\bibnamefont {Delaunay}}, \bibinfo
  {author} {\bibfnamefont {P.}~\bibnamefont {Doornenbal}}, \bibinfo {author}
  {\bibfnamefont {N.}~\bibnamefont {Fukuda}}, \bibinfo {author} {\bibfnamefont
  {J.}~\bibnamefont {Gibelin}}, \bibinfo {author} {\bibfnamefont {J.~W.}\
  \bibnamefont {Hwang}}, \bibinfo {author} {\bibfnamefont {N.}~\bibnamefont
  {Inabe}}, \bibinfo {author} {\bibfnamefont {T.}~\bibnamefont {Isobe}},
  \bibinfo {author} {\bibfnamefont {D.}~\bibnamefont {Kameda}}, \bibinfo
  {author} {\bibfnamefont {D.}~\bibnamefont {Kanno}}, \bibinfo {author}
  {\bibfnamefont {S.}~\bibnamefont {Kim}}, \bibinfo {author} {\bibfnamefont
  {N.}~\bibnamefont {Kobayashi}}, \bibinfo {author} {\bibfnamefont
  {T.}~\bibnamefont {Kobayashi}}, \bibinfo {author} {\bibfnamefont
  {T.}~\bibnamefont {Kubo}}, \bibinfo {author} {\bibfnamefont {S.}~\bibnamefont
  {Leblond}}, \bibinfo {author} {\bibfnamefont {J.}~\bibnamefont {Lee}},
  \bibinfo {author} {\bibfnamefont {F.~M.}\ \bibnamefont {Marqu\'es}}, \bibinfo
  {author} {\bibfnamefont {T.}~\bibnamefont {Motobayashi}}, \bibinfo {author}
  {\bibfnamefont {D.}~\bibnamefont {Murai}}, \bibinfo {author} {\bibfnamefont
  {T.}~\bibnamefont {Murakami}}, \bibinfo {author} {\bibfnamefont
  {K.}~\bibnamefont {Muto}}, \bibinfo {author} {\bibfnamefont {T.}~\bibnamefont
  {Nakashima}}, \bibinfo {author} {\bibfnamefont {N.}~\bibnamefont
  {Nakatsuka}}, \bibinfo {author} {\bibfnamefont {A.}~\bibnamefont {Navin}},
  \bibinfo {author} {\bibfnamefont {S.}~\bibnamefont {Nishi}}, \bibinfo
  {author} {\bibfnamefont {H.}~\bibnamefont {Otsu}}, \bibinfo {author}
  {\bibfnamefont {H.}~\bibnamefont {Sato}}, \bibinfo {author} {\bibfnamefont
  {Y.}~\bibnamefont {Satou}}, \bibinfo {author} {\bibfnamefont
  {Y.}~\bibnamefont {Shimizu}}, \bibinfo {author} {\bibfnamefont
  {H.}~\bibnamefont {Suzuki}}, \bibinfo {author} {\bibfnamefont
  {K.}~\bibnamefont {Takahashi}}, \bibinfo {author} {\bibfnamefont
  {H.}~\bibnamefont {Takeda}}, \bibinfo {author} {\bibfnamefont
  {S.}~\bibnamefont {Takeuchi}}, \bibinfo {author} {\bibfnamefont
  {Y.}~\bibnamefont {Togano}}, \bibinfo {author} {\bibfnamefont {A.~G.}\
  \bibnamefont {Tuff}}, \bibinfo {author} {\bibfnamefont {M.}~\bibnamefont
  {Vandebrouck}}, \ and\ \bibinfo {author} {\bibfnamefont {K.}~\bibnamefont
  {Yoneda}},\ }\href {\doibase 10.1103/PhysRevLett.116.102503} {\bibfield
  {journal} {\bibinfo  {journal} {Phys. Rev. Lett.}\ }\textbf {\bibinfo
  {volume} {116}},\ \bibinfo {pages} {102503} (\bibinfo {year}
  {2016})}\BibitemShut {NoStop}%
\bibitem [{\citenamefont {Revel}\ \emph {et~al.}(2018)\citenamefont {Revel},
  \citenamefont {Marqu\'es}, \citenamefont {Sorlin}, \citenamefont {Aumann},
  \citenamefont {Caesar}, \citenamefont {Holl}, \citenamefont {Panin},
  \citenamefont {Vandebrouck}, \citenamefont {Wamers}, \citenamefont
  {Alvarez-Pol}, \citenamefont {Atar}, \citenamefont {Avdeichikov},
  \citenamefont {Beceiro-Novo}, \citenamefont {Bemmerer}, \citenamefont
  {Benlliure}, \citenamefont {Bertulani}, \citenamefont {Boillos},
  \citenamefont {Boretzky}, \citenamefont {Borge}, \citenamefont {Caama\~no},
  \citenamefont {Casarejos}, \citenamefont {Catford}, \citenamefont
  {Cederk\"all}, \citenamefont {Chartier}, \citenamefont {Chulkov},
  \citenamefont {Cortina-Gil}, \citenamefont {Cravo}, \citenamefont {Crespo},
  \citenamefont {Datta~Pramanik}, \citenamefont {D\'{\i}az~Fern\'andez},
  \citenamefont {Dillmann}, \citenamefont {Elekes}, \citenamefont {Enders},
  \citenamefont {Ershova}, \citenamefont {Estrad\'e}, \citenamefont {Farinon},
  \citenamefont {Fraile}, \citenamefont {Freer}, \citenamefont {Galaviz},
  \citenamefont {Geissel}, \citenamefont {Gernh\"auser}, \citenamefont
  {Golubev}, \citenamefont {G\"obel}, \citenamefont {Hagdahl}, \citenamefont
  {Heftrich}, \citenamefont {Heil}, \citenamefont {Heine}, \citenamefont
  {Heinz}, \citenamefont {Henriques}, \citenamefont {Ignatov}, \citenamefont
  {Johansson}, \citenamefont {Jonson}, \citenamefont {Kahlbow}, \citenamefont
  {Kalantar-Nayestanaki}, \citenamefont {Kanungo}, \citenamefont {Kelic-Heil},
  \citenamefont {Knyazev}, \citenamefont {Kr\"oll}, \citenamefont {Kurz},
  \citenamefont {Labiche}, \citenamefont {Langer}, \citenamefont {Le~Bleis},
  \citenamefont {Lemmon}, \citenamefont {Lindberg}, \citenamefont {Machado},
  \citenamefont {Marganiec}, \citenamefont {Movsesyan}, \citenamefont {Nacher},
  \citenamefont {Najafi}, \citenamefont {Nilsson}, \citenamefont {Nociforo},
  \citenamefont {Paschalis}, \citenamefont {Perea}, \citenamefont {Petri},
  \citenamefont {Pietri}, \citenamefont {Plag}, \citenamefont {Reifarth},
  \citenamefont {Ribeiro}, \citenamefont {Rigollet}, \citenamefont {R\"oder},
  \citenamefont {Rossi}, \citenamefont {Savran}, \citenamefont {Scheit},
  \citenamefont {Simon}, \citenamefont {Syndikus}, \citenamefont {Taylor},
  \citenamefont {Tengblad}, \citenamefont {Thies}, \citenamefont {Togano},
  \citenamefont {Velho}, \citenamefont {Volkov}, \citenamefont {Wagner},
  \citenamefont {Weick}, \citenamefont {Wheldon}, \citenamefont {Wilson},
  \citenamefont {Winfield}, \citenamefont {Woods}, \citenamefont {Yakorev},
  \citenamefont {Zhukov}, \citenamefont {Zilges},\ and\ \citenamefont
  {Zuber}}]{PhysRevLett.120.152504}%
  \BibitemOpen
  \bibfield  {author} {\bibinfo {author} {\bibfnamefont {A.}~\bibnamefont
  {Revel}}, \bibinfo {author} {\bibfnamefont {F.~M.}\ \bibnamefont
  {Marqu\'es}}, \bibinfo {author} {\bibfnamefont {O.}~\bibnamefont {Sorlin}},
  \bibinfo {author} {\bibfnamefont {T.}~\bibnamefont {Aumann}}, \bibinfo
  {author} {\bibfnamefont {C.}~\bibnamefont {Caesar}}, \bibinfo {author}
  {\bibfnamefont {M.}~\bibnamefont {Holl}}, \bibinfo {author} {\bibfnamefont
  {V.}~\bibnamefont {Panin}}, \bibinfo {author} {\bibfnamefont
  {M.}~\bibnamefont {Vandebrouck}}, \bibinfo {author} {\bibfnamefont
  {F.}~\bibnamefont {Wamers}}, \bibinfo {author} {\bibfnamefont
  {H.}~\bibnamefont {Alvarez-Pol}}, \bibinfo {author} {\bibfnamefont
  {L.}~\bibnamefont {Atar}}, \bibinfo {author} {\bibfnamefont {V.}~\bibnamefont
  {Avdeichikov}}, \bibinfo {author} {\bibfnamefont {S.}~\bibnamefont
  {Beceiro-Novo}}, \bibinfo {author} {\bibfnamefont {D.}~\bibnamefont
  {Bemmerer}}, \bibinfo {author} {\bibfnamefont {J.}~\bibnamefont {Benlliure}},
  \bibinfo {author} {\bibfnamefont {C.~A.}\ \bibnamefont {Bertulani}}, \bibinfo
  {author} {\bibfnamefont {J.~M.}\ \bibnamefont {Boillos}}, \bibinfo {author}
  {\bibfnamefont {K.}~\bibnamefont {Boretzky}}, \bibinfo {author}
  {\bibfnamefont {M.~J.~G.}\ \bibnamefont {Borge}}, \bibinfo {author}
  {\bibfnamefont {M.}~\bibnamefont {Caama\~no}}, \bibinfo {author}
  {\bibfnamefont {E.}~\bibnamefont {Casarejos}}, \bibinfo {author}
  {\bibfnamefont {W.~N.}\ \bibnamefont {Catford}}, \bibinfo {author}
  {\bibfnamefont {J.}~\bibnamefont {Cederk\"all}}, \bibinfo {author}
  {\bibfnamefont {M.}~\bibnamefont {Chartier}}, \bibinfo {author}
  {\bibfnamefont {L.}~\bibnamefont {Chulkov}}, \bibinfo {author} {\bibfnamefont
  {D.}~\bibnamefont {Cortina-Gil}}, \bibinfo {author} {\bibfnamefont
  {E.}~\bibnamefont {Cravo}}, \bibinfo {author} {\bibfnamefont
  {R.}~\bibnamefont {Crespo}}, \bibinfo {author} {\bibfnamefont
  {U.}~\bibnamefont {Datta~Pramanik}}, \bibinfo {author} {\bibfnamefont
  {P.}~\bibnamefont {D\'{\i}az~Fern\'andez}}, \bibinfo {author} {\bibfnamefont
  {I.}~\bibnamefont {Dillmann}}, \bibinfo {author} {\bibfnamefont
  {Z.}~\bibnamefont {Elekes}}, \bibinfo {author} {\bibfnamefont
  {J.}~\bibnamefont {Enders}}, \bibinfo {author} {\bibfnamefont
  {O.}~\bibnamefont {Ershova}}, \bibinfo {author} {\bibfnamefont
  {A.}~\bibnamefont {Estrad\'e}}, \bibinfo {author} {\bibfnamefont
  {F.}~\bibnamefont {Farinon}}, \bibinfo {author} {\bibfnamefont {L.~M.}\
  \bibnamefont {Fraile}}, \bibinfo {author} {\bibfnamefont {M.}~\bibnamefont
  {Freer}}, \bibinfo {author} {\bibfnamefont {D.}~\bibnamefont {Galaviz}},
  \bibinfo {author} {\bibfnamefont {H.}~\bibnamefont {Geissel}}, \bibinfo
  {author} {\bibfnamefont {R.}~\bibnamefont {Gernh\"auser}}, \bibinfo {author}
  {\bibfnamefont {P.}~\bibnamefont {Golubev}}, \bibinfo {author} {\bibfnamefont
  {K.}~\bibnamefont {G\"obel}}, \bibinfo {author} {\bibfnamefont
  {J.}~\bibnamefont {Hagdahl}}, \bibinfo {author} {\bibfnamefont
  {T.}~\bibnamefont {Heftrich}}, \bibinfo {author} {\bibfnamefont
  {M.}~\bibnamefont {Heil}}, \bibinfo {author} {\bibfnamefont {M.}~\bibnamefont
  {Heine}}, \bibinfo {author} {\bibfnamefont {A.}~\bibnamefont {Heinz}},
  \bibinfo {author} {\bibfnamefont {A.}~\bibnamefont {Henriques}}, \bibinfo
  {author} {\bibfnamefont {A.}~\bibnamefont {Ignatov}}, \bibinfo {author}
  {\bibfnamefont {H.~T.}\ \bibnamefont {Johansson}}, \bibinfo {author}
  {\bibfnamefont {B.}~\bibnamefont {Jonson}}, \bibinfo {author} {\bibfnamefont
  {J.}~\bibnamefont {Kahlbow}}, \bibinfo {author} {\bibfnamefont
  {N.}~\bibnamefont {Kalantar-Nayestanaki}}, \bibinfo {author} {\bibfnamefont
  {R.}~\bibnamefont {Kanungo}}, \bibinfo {author} {\bibfnamefont
  {A.}~\bibnamefont {Kelic-Heil}}, \bibinfo {author} {\bibfnamefont
  {A.}~\bibnamefont {Knyazev}}, \bibinfo {author} {\bibfnamefont
  {T.}~\bibnamefont {Kr\"oll}}, \bibinfo {author} {\bibfnamefont
  {N.}~\bibnamefont {Kurz}}, \bibinfo {author} {\bibfnamefont {M.}~\bibnamefont
  {Labiche}}, \bibinfo {author} {\bibfnamefont {C.}~\bibnamefont {Langer}},
  \bibinfo {author} {\bibfnamefont {T.}~\bibnamefont {Le~Bleis}}, \bibinfo
  {author} {\bibfnamefont {R.}~\bibnamefont {Lemmon}}, \bibinfo {author}
  {\bibfnamefont {S.}~\bibnamefont {Lindberg}}, \bibinfo {author}
  {\bibfnamefont {J.}~\bibnamefont {Machado}}, \bibinfo {author} {\bibfnamefont
  {J.}~\bibnamefont {Marganiec}}, \bibinfo {author} {\bibfnamefont
  {A.}~\bibnamefont {Movsesyan}}, \bibinfo {author} {\bibfnamefont
  {E.}~\bibnamefont {Nacher}}, \bibinfo {author} {\bibfnamefont
  {M.}~\bibnamefont {Najafi}}, \bibinfo {author} {\bibfnamefont
  {T.}~\bibnamefont {Nilsson}}, \bibinfo {author} {\bibfnamefont
  {C.}~\bibnamefont {Nociforo}}, \bibinfo {author} {\bibfnamefont
  {S.}~\bibnamefont {Paschalis}}, \bibinfo {author} {\bibfnamefont
  {A.}~\bibnamefont {Perea}}, \bibinfo {author} {\bibfnamefont
  {M.}~\bibnamefont {Petri}}, \bibinfo {author} {\bibfnamefont
  {S.}~\bibnamefont {Pietri}}, \bibinfo {author} {\bibfnamefont
  {R.}~\bibnamefont {Plag}}, \bibinfo {author} {\bibfnamefont {R.}~\bibnamefont
  {Reifarth}}, \bibinfo {author} {\bibfnamefont {G.}~\bibnamefont {Ribeiro}},
  \bibinfo {author} {\bibfnamefont {C.}~\bibnamefont {Rigollet}}, \bibinfo
  {author} {\bibfnamefont {M.}~\bibnamefont {R\"oder}}, \bibinfo {author}
  {\bibfnamefont {D.}~\bibnamefont {Rossi}}, \bibinfo {author} {\bibfnamefont
  {D.}~\bibnamefont {Savran}}, \bibinfo {author} {\bibfnamefont
  {H.}~\bibnamefont {Scheit}}, \bibinfo {author} {\bibfnamefont
  {H.}~\bibnamefont {Simon}}, \bibinfo {author} {\bibfnamefont
  {I.}~\bibnamefont {Syndikus}}, \bibinfo {author} {\bibfnamefont {J.~T.}\
  \bibnamefont {Taylor}}, \bibinfo {author} {\bibfnamefont {O.}~\bibnamefont
  {Tengblad}}, \bibinfo {author} {\bibfnamefont {R.}~\bibnamefont {Thies}},
  \bibinfo {author} {\bibfnamefont {Y.}~\bibnamefont {Togano}}, \bibinfo
  {author} {\bibfnamefont {P.}~\bibnamefont {Velho}}, \bibinfo {author}
  {\bibfnamefont {V.}~\bibnamefont {Volkov}}, \bibinfo {author} {\bibfnamefont
  {A.}~\bibnamefont {Wagner}}, \bibinfo {author} {\bibfnamefont
  {H.}~\bibnamefont {Weick}}, \bibinfo {author} {\bibfnamefont
  {C.}~\bibnamefont {Wheldon}}, \bibinfo {author} {\bibfnamefont
  {G.}~\bibnamefont {Wilson}}, \bibinfo {author} {\bibfnamefont {J.~S.}\
  \bibnamefont {Winfield}}, \bibinfo {author} {\bibfnamefont {P.}~\bibnamefont
  {Woods}}, \bibinfo {author} {\bibfnamefont {D.}~\bibnamefont {Yakorev}},
  \bibinfo {author} {\bibfnamefont {M.}~\bibnamefont {Zhukov}}, \bibinfo
  {author} {\bibfnamefont {A.}~\bibnamefont {Zilges}}, \ and\ \bibinfo {author}
  {\bibfnamefont {K.}~\bibnamefont {Zuber}} (\bibinfo {collaboration}
  {${\mathrm{R}}^{3}\mathrm{B}$ Collaboration}),\ }\href {\doibase
  10.1103/PhysRevLett.120.152504} {\bibfield  {journal} {\bibinfo  {journal}
  {Phys. Rev. Lett.}\ }\textbf {\bibinfo {volume} {120}},\ \bibinfo {pages}
  {152504} (\bibinfo {year} {2018})}\BibitemShut {NoStop}%
\bibitem [{\citenamefont {Entem}\ and\ \citenamefont
  {Machleidt}(2003)}]{PhysRevC.68.041001}%
  \BibitemOpen
  \bibfield  {author} {\bibinfo {author} {\bibfnamefont {D.~R.}\ \bibnamefont
  {Entem}}\ and\ \bibinfo {author} {\bibfnamefont {R.}~\bibnamefont
  {Machleidt}},\ }\href {\doibase 10.1103/PhysRevC.68.041001} {\bibfield
  {journal} {\bibinfo  {journal} {Phys. Rev. C}\ }\textbf {\bibinfo {volume}
  {68}},\ \bibinfo {pages} {041001(R)} (\bibinfo {year} {2003})}\BibitemShut
  {NoStop}%
\bibitem [{\citenamefont {Kezerashvili}()}]{History_multi-neutron}%
  \BibitemOpen
  \bibfield  {author} {\bibinfo {author} {\bibfnamefont {R.~Y.}\ \bibnamefont
  {Kezerashvili}},\ }\href {https://arxiv.org/abs/1608.00169} {\bibinfo
  {journal} {arXiv:1608.00169 [nucl-th]}\ }\BibitemShut {NoStop}%
\bibitem [{\citenamefont {Kisamori}\ \emph {et~al.}(2016)\citenamefont
  {Kisamori}, \citenamefont {Shimoura}, \citenamefont {Miya}, \citenamefont
  {Michimasa}, \citenamefont {Ota}, \citenamefont {Assie}, \citenamefont
  {Baba}, \citenamefont {Baba}, \citenamefont {Beaumel}, \citenamefont
  {Dozono}, \citenamefont {Fujii}, \citenamefont {Fukuda}, \citenamefont {Go},
  \citenamefont {Hammache}, \citenamefont {Ideguchi}, \citenamefont {Inabe},
  \citenamefont {Itoh}, \citenamefont {Kameda}, \citenamefont {Kawase},
  \citenamefont {Kawabata}, \citenamefont {Kobayashi}, \citenamefont {Kondo},
  \citenamefont {Kubo}, \citenamefont {Kubota}, \citenamefont
  {Kurata-Nishimura}, \citenamefont {Lee}, \citenamefont {Maeda}, \citenamefont
  {Matsubara}, \citenamefont {Miki}, \citenamefont {Nishi}, \citenamefont
  {Noji}, \citenamefont {Sakaguchi}, \citenamefont {Sakai}, \citenamefont
  {Sasamoto}, \citenamefont {Sasano}, \citenamefont {Sato}, \citenamefont
  {Shimizu}, \citenamefont {Stolz}, \citenamefont {Suzuki}, \citenamefont
  {Takaki}, \citenamefont {Takeda}, \citenamefont {Takeuchi}, \citenamefont
  {Tamii}, \citenamefont {Tang}, \citenamefont {Tokieda}, \citenamefont
  {Tsumura}, \citenamefont {Uesaka}, \citenamefont {Yako}, \citenamefont
  {Yanagisawa}, \citenamefont {Yokoyama},\ and\ \citenamefont
  {Yoshida}}]{PhysRevLett.116.052501}%
  \BibitemOpen
\bibfield  {journal} {  }\bibfield  {author} {\bibinfo {author} {\bibfnamefont
  {K.}~\bibnamefont {Kisamori}}, \bibinfo {author} {\bibfnamefont
  {S.}~\bibnamefont {Shimoura}}, \bibinfo {author} {\bibfnamefont
  {H.}~\bibnamefont {Miya}}, \bibinfo {author} {\bibfnamefont {S.}~\bibnamefont
  {Michimasa}}, \bibinfo {author} {\bibfnamefont {S.}~\bibnamefont {Ota}},
  \bibinfo {author} {\bibfnamefont {M.}~\bibnamefont {Assie}}, \bibinfo
  {author} {\bibfnamefont {H.}~\bibnamefont {Baba}}, \bibinfo {author}
  {\bibfnamefont {T.}~\bibnamefont {Baba}}, \bibinfo {author} {\bibfnamefont
  {D.}~\bibnamefont {Beaumel}}, \bibinfo {author} {\bibfnamefont
  {M.}~\bibnamefont {Dozono}}, \bibinfo {author} {\bibfnamefont
  {T.}~\bibnamefont {Fujii}}, \bibinfo {author} {\bibfnamefont
  {N.}~\bibnamefont {Fukuda}}, \bibinfo {author} {\bibfnamefont
  {S.}~\bibnamefont {Go}}, \bibinfo {author} {\bibfnamefont {F.}~\bibnamefont
  {Hammache}}, \bibinfo {author} {\bibfnamefont {E.}~\bibnamefont {Ideguchi}},
  \bibinfo {author} {\bibfnamefont {N.}~\bibnamefont {Inabe}}, \bibinfo
  {author} {\bibfnamefont {M.}~\bibnamefont {Itoh}}, \bibinfo {author}
  {\bibfnamefont {D.}~\bibnamefont {Kameda}}, \bibinfo {author} {\bibfnamefont
  {S.}~\bibnamefont {Kawase}}, \bibinfo {author} {\bibfnamefont
  {T.}~\bibnamefont {Kawabata}}, \bibinfo {author} {\bibfnamefont
  {M.}~\bibnamefont {Kobayashi}}, \bibinfo {author} {\bibfnamefont
  {Y.}~\bibnamefont {Kondo}}, \bibinfo {author} {\bibfnamefont
  {T.}~\bibnamefont {Kubo}}, \bibinfo {author} {\bibfnamefont {Y.}~\bibnamefont
  {Kubota}}, \bibinfo {author} {\bibfnamefont {M.}~\bibnamefont
  {Kurata-Nishimura}}, \bibinfo {author} {\bibfnamefont {C.~S.}\ \bibnamefont
  {Lee}}, \bibinfo {author} {\bibfnamefont {Y.}~\bibnamefont {Maeda}}, \bibinfo
  {author} {\bibfnamefont {H.}~\bibnamefont {Matsubara}}, \bibinfo {author}
  {\bibfnamefont {K.}~\bibnamefont {Miki}}, \bibinfo {author} {\bibfnamefont
  {T.}~\bibnamefont {Nishi}}, \bibinfo {author} {\bibfnamefont
  {S.}~\bibnamefont {Noji}}, \bibinfo {author} {\bibfnamefont {S.}~\bibnamefont
  {Sakaguchi}}, \bibinfo {author} {\bibfnamefont {H.}~\bibnamefont {Sakai}},
  \bibinfo {author} {\bibfnamefont {Y.}~\bibnamefont {Sasamoto}}, \bibinfo
  {author} {\bibfnamefont {M.}~\bibnamefont {Sasano}}, \bibinfo {author}
  {\bibfnamefont {H.}~\bibnamefont {Sato}}, \bibinfo {author} {\bibfnamefont
  {Y.}~\bibnamefont {Shimizu}}, \bibinfo {author} {\bibfnamefont
  {A.}~\bibnamefont {Stolz}}, \bibinfo {author} {\bibfnamefont
  {H.}~\bibnamefont {Suzuki}}, \bibinfo {author} {\bibfnamefont
  {M.}~\bibnamefont {Takaki}}, \bibinfo {author} {\bibfnamefont
  {H.}~\bibnamefont {Takeda}}, \bibinfo {author} {\bibfnamefont
  {S.}~\bibnamefont {Takeuchi}}, \bibinfo {author} {\bibfnamefont
  {A.}~\bibnamefont {Tamii}}, \bibinfo {author} {\bibfnamefont
  {L.}~\bibnamefont {Tang}}, \bibinfo {author} {\bibfnamefont {H.}~\bibnamefont
  {Tokieda}}, \bibinfo {author} {\bibfnamefont {M.}~\bibnamefont {Tsumura}},
  \bibinfo {author} {\bibfnamefont {T.}~\bibnamefont {Uesaka}}, \bibinfo
  {author} {\bibfnamefont {K.}~\bibnamefont {Yako}}, \bibinfo {author}
  {\bibfnamefont {Y.}~\bibnamefont {Yanagisawa}}, \bibinfo {author}
  {\bibfnamefont {R.}~\bibnamefont {Yokoyama}}, \ and\ \bibinfo {author}
  {\bibfnamefont {K.}~\bibnamefont {Yoshida}},\ }\href {\doibase
  10.1103/PhysRevLett.116.052501} {\bibfield  {journal} {\bibinfo  {journal}
  {Phys. Rev. Lett.}\ }\textbf {\bibinfo {volume} {116}},\ \bibinfo {pages}
  {052501} (\bibinfo {year} {2016})}\BibitemShut {NoStop}%
\bibitem [{\citenamefont {Marqu\'es}\ \emph {et~al.}(2002)\citenamefont
  {Marqu\'es}, \citenamefont {Labiche}, \citenamefont {Orr}, \citenamefont
  {Ang\'elique}, \citenamefont {Axelsson}, \citenamefont {Benoit},
  \citenamefont {Bergmann}, \citenamefont {Borge}, \citenamefont {Catford},
  \citenamefont {Chappell}, \citenamefont {Clarke}, \citenamefont {Costa},
  \citenamefont {Curtis}, \citenamefont {D'Arrigo}, \citenamefont
  {de~G\'oes~Brennand}, \citenamefont {de~Oliveira~Santos}, \citenamefont
  {Dorvaux}, \citenamefont {Fazio}, \citenamefont {Freer}, \citenamefont
  {Fulton}, \citenamefont {Giardina}, \citenamefont {Gr\'evy}, \citenamefont
  {Guillemaud-Mueller}, \citenamefont {Hanappe}, \citenamefont {Heusch},
  \citenamefont {Jonson}, \citenamefont {Le~Brun}, \citenamefont {Leenhardt},
  \citenamefont {Lewitowicz}, \citenamefont {L\'opez}, \citenamefont
  {Markenroth}, \citenamefont {Mueller}, \citenamefont {Nilsson}, \citenamefont
  {Ninane}, \citenamefont {Nyman}, \citenamefont {Piqueras}, \citenamefont
  {Riisager}, \citenamefont {Laurent}, \citenamefont {Sarazin}, \citenamefont
  {Singer}, \citenamefont {Sorlin},\ and\ \citenamefont
  {Stuttg\'e}}]{PhysRevC.65.044006}%
  \BibitemOpen
  \bibfield  {author} {\bibinfo {author} {\bibfnamefont {F.~M.}\ \bibnamefont
  {Marqu\'es}}, \bibinfo {author} {\bibfnamefont {M.}~\bibnamefont {Labiche}},
  \bibinfo {author} {\bibfnamefont {N.~A.}\ \bibnamefont {Orr}}, \bibinfo
  {author} {\bibfnamefont {J.~C.}\ \bibnamefont {Ang\'elique}}, \bibinfo
  {author} {\bibfnamefont {L.}~\bibnamefont {Axelsson}}, \bibinfo {author}
  {\bibfnamefont {B.}~\bibnamefont {Benoit}}, \bibinfo {author} {\bibfnamefont
  {U.~C.}\ \bibnamefont {Bergmann}}, \bibinfo {author} {\bibfnamefont
  {M.~J.~G.}\ \bibnamefont {Borge}}, \bibinfo {author} {\bibfnamefont {W.~N.}\
  \bibnamefont {Catford}}, \bibinfo {author} {\bibfnamefont {S.~P.~G.}\
  \bibnamefont {Chappell}}, \bibinfo {author} {\bibfnamefont {N.~M.}\
  \bibnamefont {Clarke}}, \bibinfo {author} {\bibfnamefont {G.}~\bibnamefont
  {Costa}}, \bibinfo {author} {\bibfnamefont {N.}~\bibnamefont {Curtis}},
  \bibinfo {author} {\bibfnamefont {A.}~\bibnamefont {D'Arrigo}}, \bibinfo
  {author} {\bibfnamefont {E.}~\bibnamefont {de~G\'oes~Brennand}}, \bibinfo
  {author} {\bibfnamefont {F.}~\bibnamefont {de~Oliveira~Santos}}, \bibinfo
  {author} {\bibfnamefont {O.}~\bibnamefont {Dorvaux}}, \bibinfo {author}
  {\bibfnamefont {G.}~\bibnamefont {Fazio}}, \bibinfo {author} {\bibfnamefont
  {M.}~\bibnamefont {Freer}}, \bibinfo {author} {\bibfnamefont {B.~R.}\
  \bibnamefont {Fulton}}, \bibinfo {author} {\bibfnamefont {G.}~\bibnamefont
  {Giardina}}, \bibinfo {author} {\bibfnamefont {S.}~\bibnamefont {Gr\'evy}},
  \bibinfo {author} {\bibfnamefont {D.}~\bibnamefont {Guillemaud-Mueller}},
  \bibinfo {author} {\bibfnamefont {F.}~\bibnamefont {Hanappe}}, \bibinfo
  {author} {\bibfnamefont {B.}~\bibnamefont {Heusch}}, \bibinfo {author}
  {\bibfnamefont {B.}~\bibnamefont {Jonson}}, \bibinfo {author} {\bibfnamefont
  {C.}~\bibnamefont {Le~Brun}}, \bibinfo {author} {\bibfnamefont
  {S.}~\bibnamefont {Leenhardt}}, \bibinfo {author} {\bibfnamefont
  {M.}~\bibnamefont {Lewitowicz}}, \bibinfo {author} {\bibfnamefont {M.~J.}\
  \bibnamefont {L\'opez}}, \bibinfo {author} {\bibfnamefont {K.}~\bibnamefont
  {Markenroth}}, \bibinfo {author} {\bibfnamefont {A.~C.}\ \bibnamefont
  {Mueller}}, \bibinfo {author} {\bibfnamefont {T.}~\bibnamefont {Nilsson}},
  \bibinfo {author} {\bibfnamefont {A.}~\bibnamefont {Ninane}}, \bibinfo
  {author} {\bibfnamefont {G.}~\bibnamefont {Nyman}}, \bibinfo {author}
  {\bibfnamefont {I.}~\bibnamefont {Piqueras}}, \bibinfo {author}
  {\bibfnamefont {K.}~\bibnamefont {Riisager}}, \bibinfo {author}
  {\bibfnamefont {M.~G.~S.}\ \bibnamefont {Laurent}}, \bibinfo {author}
  {\bibfnamefont {F.}~\bibnamefont {Sarazin}}, \bibinfo {author} {\bibfnamefont
  {S.~M.}\ \bibnamefont {Singer}}, \bibinfo {author} {\bibfnamefont
  {O.}~\bibnamefont {Sorlin}}, \ and\ \bibinfo {author} {\bibfnamefont
  {L.}~\bibnamefont {Stuttg\'e}},\ }\href {\doibase 10.1103/PhysRevC.65.044006}
  {\bibfield  {journal} {\bibinfo  {journal} {Phys. Rev. C}\ }\textbf {\bibinfo
  {volume} {65}},\ \bibinfo {pages} {044006} (\bibinfo {year}
  {2002})}\BibitemShut {NoStop}%
\bibitem [{\citenamefont {Pieper}(2003)}]{PhysRevLett.90.252501}%
  \BibitemOpen
  \bibfield  {author} {\bibinfo {author} {\bibfnamefont {S.~C.}\ \bibnamefont
  {Pieper}},\ }\href {\doibase 10.1103/PhysRevLett.90.252501} {\bibfield
  {journal} {\bibinfo  {journal} {Phys. Rev. Lett.}\ }\textbf {\bibinfo
  {volume} {90}},\ \bibinfo {pages} {252501} (\bibinfo {year}
  {2003})}\BibitemShut {NoStop}%
\bibitem [{\citenamefont {Bertulani}\ and\ \citenamefont
  {Zelevinsky}(2003)}]{Bertulani_2003}%
  \BibitemOpen
  \bibfield  {author} {\bibinfo {author} {\bibfnamefont {C.~A.}\ \bibnamefont
  {Bertulani}}\ and\ \bibinfo {author} {\bibfnamefont {V.}~\bibnamefont
  {Zelevinsky}},\ }\href {\doibase 10.1088/0954-3899/29/10/309} {\bibfield
  {journal} {\bibinfo  {journal} {J. Phys. G. Nucl. Part. Phys}\ }\textbf
  {\bibinfo {volume} {29}},\ \bibinfo {pages} {2431} (\bibinfo {year}
  {2003})}\BibitemShut {NoStop}%
\bibitem [{\citenamefont {Paschalis}\ \emph {et~al.}()\citenamefont {Paschalis}
  \emph {et~al.}}]{tetra_neutron_1}%
  \BibitemOpen
  \bibfield  {author} {\bibinfo {author} {\bibfnamefont {S.}~\bibnamefont
  {Paschalis}} \emph {et~al.},\ }\href@noop {} {\bibinfo  {journal} {Report No.
  NP1406-SAMURAI19}\ }\BibitemShut {NoStop}%
\bibitem [{\citenamefont {Kisamori}\ \emph {et~al.}()\citenamefont {Kisamori}
  \emph {et~al.}}]{tetra_neutron_2}%
  \BibitemOpen
\bibfield  {journal} {  }\bibfield  {author} {\bibinfo {author} {\bibfnamefont
  {K.}~\bibnamefont {Kisamori}} \emph {et~al.},\ }\href@noop {} {\bibinfo
  {journal} {Report No. NP1512-SAMURAI34}\ }\BibitemShut {NoStop}%
\bibitem [{\citenamefont {Shimoura}\ \emph {et~al.}()\citenamefont {Shimoura}
  \emph {et~al.}}]{tetra_neutron_3}%
  \BibitemOpen
\bibfield  {journal} {  }\bibfield  {author} {\bibinfo {author} {\bibfnamefont
  {S.}~\bibnamefont {Shimoura}} \emph {et~al.},\ }\href@noop {} {\bibinfo
  {journal} {Report No. NP1512-SHARAQ10}\ }\BibitemShut {NoStop}%
\bibitem [{\citenamefont {Shirokov}\ \emph {et~al.}(2016)\citenamefont
  {Shirokov}, \citenamefont {Papadimitriou}, \citenamefont {Mazur},
  \citenamefont {Mazur}, \citenamefont {Roth},\ and\ \citenamefont
  {Vary}}]{PhysRevLett.117.182502}%
  \BibitemOpen
\bibfield  {journal} {  }\bibfield  {author} {\bibinfo {author} {\bibfnamefont
  {A.~M.}\ \bibnamefont {Shirokov}}, \bibinfo {author} {\bibfnamefont
  {G.}~\bibnamefont {Papadimitriou}}, \bibinfo {author} {\bibfnamefont {A.~I.}\
  \bibnamefont {Mazur}}, \bibinfo {author} {\bibfnamefont {I.~A.}\ \bibnamefont
  {Mazur}}, \bibinfo {author} {\bibfnamefont {R.}~\bibnamefont {Roth}}, \ and\
  \bibinfo {author} {\bibfnamefont {J.~P.}\ \bibnamefont {Vary}},\ }\href
  {\doibase 10.1103/PhysRevLett.117.182502} {\bibfield  {journal} {\bibinfo
  {journal} {Phys. Rev. Lett.}\ }\textbf {\bibinfo {volume} {117}},\ \bibinfo
  {pages} {182502} (\bibinfo {year} {2016})}\BibitemShut {NoStop}%
\bibitem [{\citenamefont {Gandolfi}\ \emph {et~al.}(2017)\citenamefont
  {Gandolfi}, \citenamefont {Hammer}, \citenamefont {Klos}, \citenamefont
  {Lynn},\ and\ \citenamefont {Schwenk}}]{PhysRevLett.118.232501}%
  \BibitemOpen
  \bibfield  {author} {\bibinfo {author} {\bibfnamefont {S.}~\bibnamefont
  {Gandolfi}}, \bibinfo {author} {\bibfnamefont {H.-W.}\ \bibnamefont
  {Hammer}}, \bibinfo {author} {\bibfnamefont {P.}~\bibnamefont {Klos}},
  \bibinfo {author} {\bibfnamefont {J.~E.}\ \bibnamefont {Lynn}}, \ and\
  \bibinfo {author} {\bibfnamefont {A.}~\bibnamefont {Schwenk}},\ }\href
  {\doibase 10.1103/PhysRevLett.118.232501} {\bibfield  {journal} {\bibinfo
  {journal} {Phys. Rev. Lett.}\ }\textbf {\bibinfo {volume} {118}},\ \bibinfo
  {pages} {232501} (\bibinfo {year} {2017})}\BibitemShut {NoStop}%
\bibitem [{\citenamefont {Fossez}\ \emph {et~al.}(2017)\citenamefont {Fossez},
  \citenamefont {Rotureau}, \citenamefont {Michel},\ and\ \citenamefont
  {P\l{}oszajczak}}]{PhysRevLett.119.032501}%
  \BibitemOpen
  \bibfield  {author} {\bibinfo {author} {\bibfnamefont {K.}~\bibnamefont
  {Fossez}}, \bibinfo {author} {\bibfnamefont {J.}~\bibnamefont {Rotureau}},
  \bibinfo {author} {\bibfnamefont {N.}~\bibnamefont {Michel}}, \ and\ \bibinfo
  {author} {\bibfnamefont {M.}~\bibnamefont {P\l{}oszajczak}},\ }\href
  {\doibase 10.1103/PhysRevLett.119.032501} {\bibfield  {journal} {\bibinfo
  {journal} {Phys. Rev. Lett.}\ }\textbf {\bibinfo {volume} {119}},\ \bibinfo
  {pages} {032501} (\bibinfo {year} {2017})}\BibitemShut {NoStop}%
\bibitem [{\citenamefont {Deltuva}(2018{\natexlab{a}})}]{PhysRevC.97.034001}%
  \BibitemOpen
  \bibfield  {author} {\bibinfo {author} {\bibfnamefont {A.}~\bibnamefont
  {Deltuva}},\ }\href {\doibase 10.1103/PhysRevC.97.034001} {\bibfield
  {journal} {\bibinfo  {journal} {Phys. Rev. C}\ }\textbf {\bibinfo {volume}
  {97}},\ \bibinfo {pages} {034001} (\bibinfo {year}
  {2018}{\natexlab{a}})}\BibitemShut {NoStop}%
\bibitem [{\citenamefont {Deltuva}(2018{\natexlab{b}})}]{DELTUVA2018238}%
  \BibitemOpen
  \bibfield  {author} {\bibinfo {author} {\bibfnamefont {A.}~\bibnamefont
  {Deltuva}},\ }\href {\doibase https://doi.org/10.1016/j.physletb.2018.05.041}
  {\bibfield  {journal} {\bibinfo  {journal} {Phys. Lett. B}\ }\textbf
  {\bibinfo {volume} {782}},\ \bibinfo {pages} {238 } (\bibinfo {year}
  {2018}{\natexlab{b}})}\BibitemShut {NoStop}%
\bibitem [{\citenamefont {Lazauskas}\ and\ \citenamefont
  {Carbonell}(2005)}]{PhysRevC.72.034003}%
  \BibitemOpen
  \bibfield  {author} {\bibinfo {author} {\bibfnamefont {R.}~\bibnamefont
  {Lazauskas}}\ and\ \bibinfo {author} {\bibfnamefont {J.}~\bibnamefont
  {Carbonell}},\ }\href {\doibase 10.1103/PhysRevC.72.034003} {\bibfield
  {journal} {\bibinfo  {journal} {Phys. Rev. C}\ }\textbf {\bibinfo {volume}
  {72}},\ \bibinfo {pages} {034003} (\bibinfo {year} {2005})}\BibitemShut
  {NoStop}%
\bibitem [{\citenamefont {Hiyama}\ \emph {et~al.}(2016)\citenamefont {Hiyama},
  \citenamefont {Lazauskas}, \citenamefont {Carbonell},\ and\ \citenamefont
  {Kamimura}}]{PhysRevC.93.044004}%
  \BibitemOpen
  \bibfield  {author} {\bibinfo {author} {\bibfnamefont {E.}~\bibnamefont
  {Hiyama}}, \bibinfo {author} {\bibfnamefont {R.}~\bibnamefont {Lazauskas}},
  \bibinfo {author} {\bibfnamefont {J.}~\bibnamefont {Carbonell}}, \ and\
  \bibinfo {author} {\bibfnamefont {M.}~\bibnamefont {Kamimura}},\ }\href
  {\doibase 10.1103/PhysRevC.93.044004} {\bibfield  {journal} {\bibinfo
  {journal} {Phys. Rev. C}\ }\textbf {\bibinfo {volume} {93}},\ \bibinfo
  {pages} {044004} (\bibinfo {year} {2016})}\BibitemShut {NoStop}%
\bibitem [{\citenamefont {Shirokov}()}]{Tetraneutron_NCSM2018}%
  \BibitemOpen
  \bibfield  {author} {\bibinfo {author} {\bibfnamefont {A.~M.}\ \bibnamefont
  {Shirokov}},\ }\href@noop {} {\bibinfo  {journal} {private communication}\
  }\BibitemShut {NoStop}%
\bibitem [{\citenamefont {Gandolfi}\ \emph {et~al.}(2019)\citenamefont
  {Gandolfi}, \citenamefont {Hammer}, \citenamefont {Klos}, \citenamefont
  {Lynn},\ and\ \citenamefont {Schwenk}}]{PhysRevLett.123.069202}%
  \BibitemOpen
\bibfield  {journal} {  }\bibfield  {author} {\bibinfo {author} {\bibfnamefont
  {S.}~\bibnamefont {Gandolfi}}, \bibinfo {author} {\bibfnamefont {H.-W.}\
  \bibnamefont {Hammer}}, \bibinfo {author} {\bibfnamefont {P.}~\bibnamefont
  {Klos}}, \bibinfo {author} {\bibfnamefont {J.~E.}\ \bibnamefont {Lynn}}, \
  and\ \bibinfo {author} {\bibfnamefont {A.}~\bibnamefont {Schwenk}},\ }\href
  {\doibase 10.1103/PhysRevLett.123.069202} {\bibfield  {journal} {\bibinfo
  {journal} {Phys. Rev. Lett.}\ }\textbf {\bibinfo {volume} {123}},\ \bibinfo
  {pages} {069202} (\bibinfo {year} {2019})}\BibitemShut {NoStop}%
\bibitem [{\citenamefont {Papadimitriou}\ \emph {et~al.}(2013)\citenamefont
  {Papadimitriou}, \citenamefont {Rotureau}, \citenamefont {Michel},
  \citenamefont {P\l{}oszajczak},\ and\ \citenamefont
  {Barrett}}]{PhysRevC.88.044318}%
  \BibitemOpen
  \bibfield  {author} {\bibinfo {author} {\bibfnamefont {G.}~\bibnamefont
  {Papadimitriou}}, \bibinfo {author} {\bibfnamefont {J.}~\bibnamefont
  {Rotureau}}, \bibinfo {author} {\bibfnamefont {N.}~\bibnamefont {Michel}},
  \bibinfo {author} {\bibfnamefont {M.}~\bibnamefont {P\l{}oszajczak}}, \ and\
  \bibinfo {author} {\bibfnamefont {B.~R.}\ \bibnamefont {Barrett}},\ }\href
  {\doibase 10.1103/PhysRevC.88.044318} {\bibfield  {journal} {\bibinfo
  {journal} {Phys. Rev. C}\ }\textbf {\bibinfo {volume} {88}},\ \bibinfo
  {pages} {044318} (\bibinfo {year} {2013})}\BibitemShut {NoStop}%
\bibitem [{\citenamefont {Berggren}(1968)}]{BERGGREN1968265}%
  \BibitemOpen
  \bibfield  {author} {\bibinfo {author} {\bibfnamefont {T.}~\bibnamefont
  {Berggren}},\ }\href {\doibase https://doi.org/10.1016/0375-9474(68)90593-9}
  {\bibfield  {journal} {\bibinfo  {journal} {Nucl. Phys. A}\ }\textbf
  {\bibinfo {volume} {109}},\ \bibinfo {pages} {265 } (\bibinfo {year}
  {1968})}\BibitemShut {NoStop}%
\bibitem [{\citenamefont {Volya}(2009)}]{PhysRevC.79.044308}%
  \BibitemOpen
  \bibfield  {author} {\bibinfo {author} {\bibfnamefont {A.}~\bibnamefont
  {Volya}},\ }\href {\doibase 10.1103/PhysRevC.79.044308} {\bibfield  {journal}
  {\bibinfo  {journal} {Phys. Rev. C}\ }\textbf {\bibinfo {volume} {79}},\
  \bibinfo {pages} {044308} (\bibinfo {year} {2009})}\BibitemShut {NoStop}%
\bibitem [{\citenamefont {Michel}\ \emph {et~al.}(2002)\citenamefont {Michel},
  \citenamefont {Nazarewicz}, \citenamefont {P\l{}oszajczak},\ and\
  \citenamefont {Bennaceur}}]{PhysRevLett.89.042502}%
  \BibitemOpen
  \bibfield  {author} {\bibinfo {author} {\bibfnamefont {N.}~\bibnamefont
  {Michel}}, \bibinfo {author} {\bibfnamefont {W.}~\bibnamefont {Nazarewicz}},
  \bibinfo {author} {\bibfnamefont {M.}~\bibnamefont {P\l{}oszajczak}}, \ and\
  \bibinfo {author} {\bibfnamefont {K.}~\bibnamefont {Bennaceur}},\ }\href
  {\doibase 10.1103/PhysRevLett.89.042502} {\bibfield  {journal} {\bibinfo
  {journal} {Phys. Rev. Lett.}\ }\textbf {\bibinfo {volume} {89}},\ \bibinfo
  {pages} {042502} (\bibinfo {year} {2002})}\BibitemShut {NoStop}%
\bibitem [{\citenamefont {Id~Betan}\ \emph {et~al.}(2002)\citenamefont
  {Id~Betan}, \citenamefont {Liotta}, \citenamefont {Sandulescu},\ and\
  \citenamefont {Vertse}}]{PhysRevLett.89.042501}%
  \BibitemOpen
  \bibfield  {author} {\bibinfo {author} {\bibfnamefont {R.}~\bibnamefont
  {Id~Betan}}, \bibinfo {author} {\bibfnamefont {R.~J.}\ \bibnamefont
  {Liotta}}, \bibinfo {author} {\bibfnamefont {N.}~\bibnamefont {Sandulescu}},
  \ and\ \bibinfo {author} {\bibfnamefont {T.}~\bibnamefont {Vertse}},\ }\href
  {\doibase 10.1103/PhysRevLett.89.042501} {\bibfield  {journal} {\bibinfo
  {journal} {Phys. Rev. Lett.}\ }\textbf {\bibinfo {volume} {89}},\ \bibinfo
  {pages} {042501} (\bibinfo {year} {2002})}\BibitemShut {NoStop}%
\bibitem [{\citenamefont {Barrett}\ \emph {et~al.}(2013)\citenamefont
  {Barrett}, \citenamefont {Navr\'atil},\ and\ \citenamefont
  {Vary}}]{BARRETT2013131}%
  \BibitemOpen
  \bibfield  {author} {\bibinfo {author} {\bibfnamefont {B.~R.}\ \bibnamefont
  {Barrett}}, \bibinfo {author} {\bibfnamefont {P.}~\bibnamefont {Navr\'atil}},
  \ and\ \bibinfo {author} {\bibfnamefont {J.~P.}\ \bibnamefont {Vary}},\
  }\href {\doibase https://doi.org/10.1016/j.ppnp.2012.10.003} {\bibfield
  {journal} {\bibinfo  {journal} {Prog. Part. Nucl. Phys.}\ }\textbf {\bibinfo
  {volume} {69}},\ \bibinfo {pages} {131 } (\bibinfo {year}
  {2013})}\BibitemShut {NoStop}%
\bibitem [{\citenamefont {Michel}\ \emph {et~al.}(2008)\citenamefont {Michel},
  \citenamefont {Nazarewicz}, \citenamefont {P{\l}oszajczak},\ and\
  \citenamefont {Vertse}}]{Michel_2008}%
  \BibitemOpen
  \bibfield  {author} {\bibinfo {author} {\bibfnamefont {N.}~\bibnamefont
  {Michel}}, \bibinfo {author} {\bibfnamefont {W.}~\bibnamefont {Nazarewicz}},
  \bibinfo {author} {\bibfnamefont {M.}~\bibnamefont {P{\l}oszajczak}}, \ and\
  \bibinfo {author} {\bibfnamefont {T.}~\bibnamefont {Vertse}},\ }\href
  {\doibase 10.1088/0954-3899/36/1/013101} {\bibfield  {journal} {\bibinfo
  {journal} {J. Phys. G. Nucl. Part. Phys}\ }\textbf {\bibinfo {volume} {36}},\
  \bibinfo {pages} {013101} (\bibinfo {year} {2008})}\BibitemShut {NoStop}%
\bibitem [{\citenamefont {Romero-Redondo}\ \emph {et~al.}(2016)\citenamefont
  {Romero-Redondo}, \citenamefont {Quaglioni}, \citenamefont {Navr\'atil},\
  and\ \citenamefont {Hupin}}]{PhysRevLett.117.222501}%
  \BibitemOpen
  \bibfield  {author} {\bibinfo {author} {\bibfnamefont {C.}~\bibnamefont
  {Romero-Redondo}}, \bibinfo {author} {\bibfnamefont {S.}~\bibnamefont
  {Quaglioni}}, \bibinfo {author} {\bibfnamefont {P.}~\bibnamefont
  {Navr\'atil}}, \ and\ \bibinfo {author} {\bibfnamefont {G.}~\bibnamefont
  {Hupin}},\ }\href {\doibase 10.1103/PhysRevLett.117.222501} {\bibfield
  {journal} {\bibinfo  {journal} {Phys. Rev. Lett.}\ }\textbf {\bibinfo
  {volume} {117}},\ \bibinfo {pages} {222501} (\bibinfo {year}
  {2016})}\BibitemShut {NoStop}%
\bibitem [{\citenamefont {Sun}\ \emph {et~al.}(2017)\citenamefont {Sun},
  \citenamefont {Wu}, \citenamefont {Zhao}, \citenamefont {Hu}, \citenamefont
  {Dai},\ and\ \citenamefont {Xu}}]{SUN2017227}%
  \BibitemOpen
  \bibfield  {author} {\bibinfo {author} {\bibfnamefont {Z.}~\bibnamefont
  {Sun}}, \bibinfo {author} {\bibfnamefont {Q.}~\bibnamefont {Wu}}, \bibinfo
  {author} {\bibfnamefont {Z.}~\bibnamefont {Zhao}}, \bibinfo {author}
  {\bibfnamefont {B.}~\bibnamefont {Hu}}, \bibinfo {author} {\bibfnamefont
  {S.}~\bibnamefont {Dai}}, \ and\ \bibinfo {author} {\bibfnamefont
  {F.}~\bibnamefont {Xu}},\ }\href {\doibase
  https://doi.org/10.1016/j.physletb.2017.03.054} {\bibfield  {journal}
  {\bibinfo  {journal} {Phys. Lett. B}\ }\textbf {\bibinfo {volume} {769}},\
  \bibinfo {pages} {227 } (\bibinfo {year} {2017})}\BibitemShut {NoStop}%
\bibitem [{\citenamefont {Hagen}\ \emph {et~al.}(2012)\citenamefont {Hagen},
  \citenamefont {Hjorth-Jensen}, \citenamefont {Jansen}, \citenamefont
  {Machleidt},\ and\ \citenamefont {Papenbrock}}]{PhysRevLett.108.242501}%
  \BibitemOpen
  \bibfield  {author} {\bibinfo {author} {\bibfnamefont {G.}~\bibnamefont
  {Hagen}}, \bibinfo {author} {\bibfnamefont {M.}~\bibnamefont
  {Hjorth-Jensen}}, \bibinfo {author} {\bibfnamefont {G.~R.}\ \bibnamefont
  {Jansen}}, \bibinfo {author} {\bibfnamefont {R.}~\bibnamefont {Machleidt}}, \
  and\ \bibinfo {author} {\bibfnamefont {T.}~\bibnamefont {Papenbrock}},\
  }\href {\doibase 10.1103/PhysRevLett.108.242501} {\bibfield  {journal}
  {\bibinfo  {journal} {Phys. Rev. Lett.}\ }\textbf {\bibinfo {volume} {108}},\
  \bibinfo {pages} {242501} (\bibinfo {year} {2012})}\BibitemShut {NoStop}%
\end{thebibliography}%

\end{document}